\documentclass[aps,prl,groupedaddress,showpacs,tighten,nofootinbib,preprint]
{revtex4}

\usepackage{graphicx,amssymb,amsmath}

\usepackage{times}

\begin{document}

\preprint{FERMILAB-PUB-06-208-E}

\title{Deep shower interpretation of the cosmic ray events observed in excess of the Greisen-Zatsepin-Kuzmin energy}

\author{Aaron S. Chou}

\affiliation{Fermi National Accelerator Laboratory, 
	PO Box 500, Batavia, IL  60510-0500}


\begin{abstract}

We consider the possibility that the ultra-high-energy cosmic ray flux has a small component of exotic particles which create showers much deeper in the atmosphere than ordinary hadronic primaries.  It is shown that applying the conventional AGASA/HiRes/Auger data analysis procedures to such exotic events results in large systematic biases in the energy spectrum measurement which may distort the shape of the measured spectrum near the expected Greisen-Zatsepin-Kuzmin (GZK) cutoff energy.  Sub-GZK exotic showers may be mis-reconstructed with much higher energies and mimic super-GZK events.  Alternatively, super-GZK exotic showers may elude detection by conventional fluorescence analysis techniques. 

\end{abstract}

\pacs{95.55.Vj;96.50.sb;98.70.Sa\hspace{0.5cm} FERMILAB-PUB-06-208-E}

\maketitle

\section{Introduction}

The measurement by the AGASA ground array \cite{Takeda:1998ps} of an unabated ultra-high-energy cosmic ray (UHECR) flux beyond the Greisen-Zatsepin-Kuzmin (GZK) energy of $\sim 5\times 10^{19}$ eV \cite{Greisen:1966jv} \cite{Zatsepin:1966} has provoked much speculation in the particle astrophysics community.  Meanwhile, the HiRes fluorescence experiment has reported evidence of the expected GZK suppression \cite{Abbasi:2002s} \cite{Abbasi:2005s}.  Explanations of the AGASA super-GZK events have mainly focused on mechanisms by which the energetic particles may evade interaction with the CMB.  If the UHECRs are ordinary hadronic particles, then their sources must be very close, within $\sim$100 Mpc.  No local astrophysical sources have been compellingly identified, though interesting possibilities do exist \cite{Farrar:2005me}.  In some models, local decays of heavy cosmological relic particles create the UHECR flux.  In this case, the flux of conventional particles is usually expected to be dominated by energetic photons from $\pi^0$ decay.  These types of models are starting to become constrained by limits on the photon fraction of the UHECR flux \cite{Gelmini:2005wu} \cite{Risse:2005jr} \cite{Risse:2005hi} \cite{Rubtsov:2006tt}.

Alternatively if the UHECRs originate very far away, then in order to propagate, their interactions with the CMB must be greatly suppressed.  If, for example, Lorentz invariance is violated for particles with large boosts relative to our local cosmological rest frame, then the thresholds for the various interactions and hence the GZK feature can be shifted towards higher energies \cite{Kirzhnits:1972sg} \cite{Sato:1972wy} \cite{Gonzalez-Mestres:1997mq} \cite{Gonzalez-Mestres:1997zg} \cite{Coleman:1998ti} \cite{Stecker:2001vb}.  Other conventional particles with small interaction cross-sections include neutrinos and axions.  Neutrinos could induce hadronic air showers if their hadronic interaction cross-sections were greatly enhanced due to new physics at the TeV scale\cite{Feng:2001ib} \cite{Anchordoqui:2001cg} \cite{Ahn:2003qn} \cite{Anchordoqui:2005ey}.  Axion-like particles could propagate large distances before converting locally to photons which then shower in the atmosphere \cite{Gorbunov:2001gc}.  Of course, one can always postulate the existence of a new species of particle with interaction processes tuned to solve the propagation problem and to still make detectable air showers.  For example, the GZK energy threshold for a heavy particle of mass $M_X$ is increased by a factor of $M_X/M_{proton}$.  To create detectable air showers which can mimic ordinary hadronic events at moderate zenith angles, the cross-section $\sigma_{X-nucleon}$ would need to be in the range 10-1000 mb.  Specific models for such ``UHEcrons'' were analyzed in \cite{Chung:1997rz} \cite{Albuquerque:1998va} \cite{Berezinsky:2001fy} \cite{Kachelriess:2003yy}.  Some recent experimental limits from CDMS and Edelweiss data on heavy particle fluxes were reported in \cite{Albuquerque:2003ei}.  Also, searches for neutrino showers at large zenith angles place limits on the flux of very deeply penetrating particles with cross-sections much smaller than we are considering in this work \cite{Anchordoqui:2003jr} \cite{Baltrusaitis:1985mt}.

Regardless of the mechanism by which the super-GZK cosmic rays evade interaction with the CMB, the properties of the air showers produced at the Earth may be quite model-dependent.  However, all current experiments analyze their air shower data with the assumption that the primary particles are ordinary hadrons.  There is no guarantee that analyzing exotic events in this way would yield sensible results.  For example, exotic events which look very different from conventional air showers could be rejected by event selection criteria designed to select precisely those events which look like ordinary air showers and not like noise.  In some cases, the data from an exotic shower could be mis-reconstructed to look somewhat like a ordinary hadronic shower.  In this case, even though the event would pass the event selection cuts, the reconstructed quantities such as the energy and arrival direction would very likely not be reliable.  In the next three sections, we will discuss how exotic events would be treated by the detectors and the data analysis techniques of the AGASA, HiRes, and Auger experiments.  We will focus specifically on the case where exotic cosmic rays with small cross-sections $\sigma_{X-nucleon}$ produce deeply-penetrating air showers at moderate zenith angles.

\section{Deep showers as viewed by AGASA}

AGASA is a surface detector array of scintillators which measure the air shower particle flux at 900 m above sea level, the equivalent of 920 g/cm$^2$ of vertical atmospheric depth.  Hadronic showers typically reach their maximum particle flux at a depth of $X_{max}\sim$ 750 g/cm$^2$, at which ionization loss and charged pion decay curtail the further growth of the shower.  The value of $X_{max}$ is determined by both the hadronic interaction length and interaction dynamics of the primary cosmic ray, and the interaction/radiation lengths of the daughter particles produced in the shower.   For hadronic primaries, it is expected that the energy of the primary is efficiently transported to the electromagnetic portion of the shower, which then develops in a predictable ``universal'' manner, independent of the initial source of the energy \cite{Chou:2005yq}.  This is the principle upon which calorimetric measurements of the air showers via the electromagnetic particle flux is based.  Whereas a fluorescence telescope like HiRes images the longitudinal development of the electromagnetic shower, a surface detector array like AGASA samples the flux as it hits the ground and then estimates the total energy based on the magnitude of the intercepted flux.

Because of variations typically of order $\pm$50 g/cm$^2$ on the position of $X_{max}$ from shower to shower due to statistical fluctuations in the first few hadronic interactions ($\lambda_I \sim$ 70 g/cm$^2$), the magnitude of the total ground level flux can vary even for showers at a fixed energy.  Since the flux attenuation length $\sim$70 g/cm$^2$ due to ionization loss is about the same magnitude as the $X_{max}$ fluctuations, measuring or estimating the total particle flux at ground would give quite poor energy resolution.  Instead, the estimated particle flux $\rho(R)$ at some finite transverse core distance $R$ is found to be much better correlated with the shower energy \cite{Hillas:1971}.  If a shower happens to be closer to the ground, then the total ground flux increases but not all of the increased flux has enough time to transversely migrate to the detection position $R$ before hitting the ground.  Similarly, if a shower is positioned further from the ground, the overall ground flux is decreased but more of that flux can reach large distances from the core.  The AGASA array is designed with a detector spacing of 1 km in order to optimize the reconstruction of the flux at $R=$ 600 m core distance, and $\rho(600)$ is the parameter they estimate from their data in order to obtain the energy measurement.

\begin{figure}
\begin{minipage}[t]{0.48\textwidth}
\mbox{}\\
\centerline{\includegraphics[width=\textwidth]{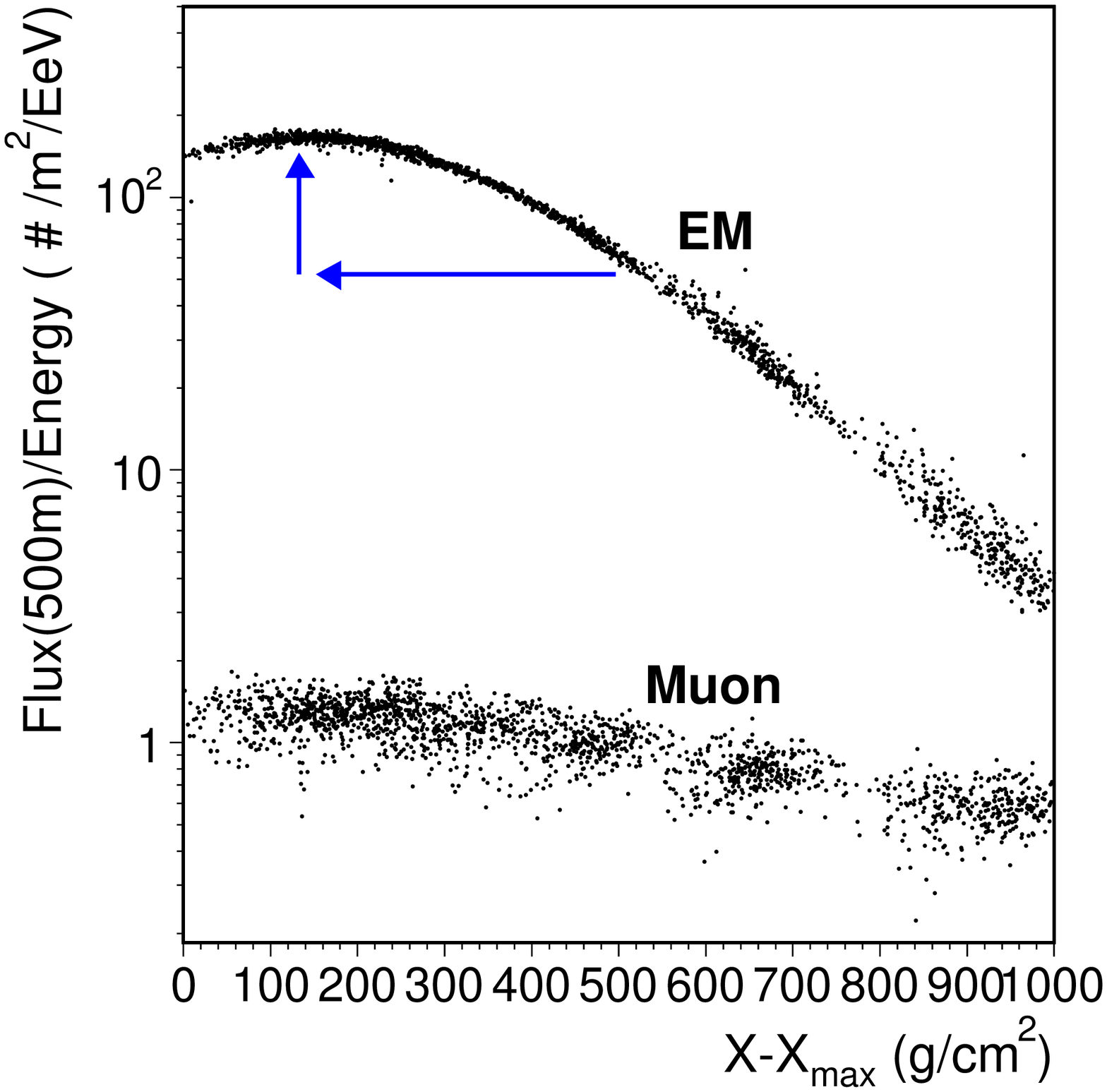}}
\end{minipage}
\hfill
\begin{minipage}[t]{0.48\textwidth}
\mbox{}\\
\centerline{\includegraphics[width=\textwidth]{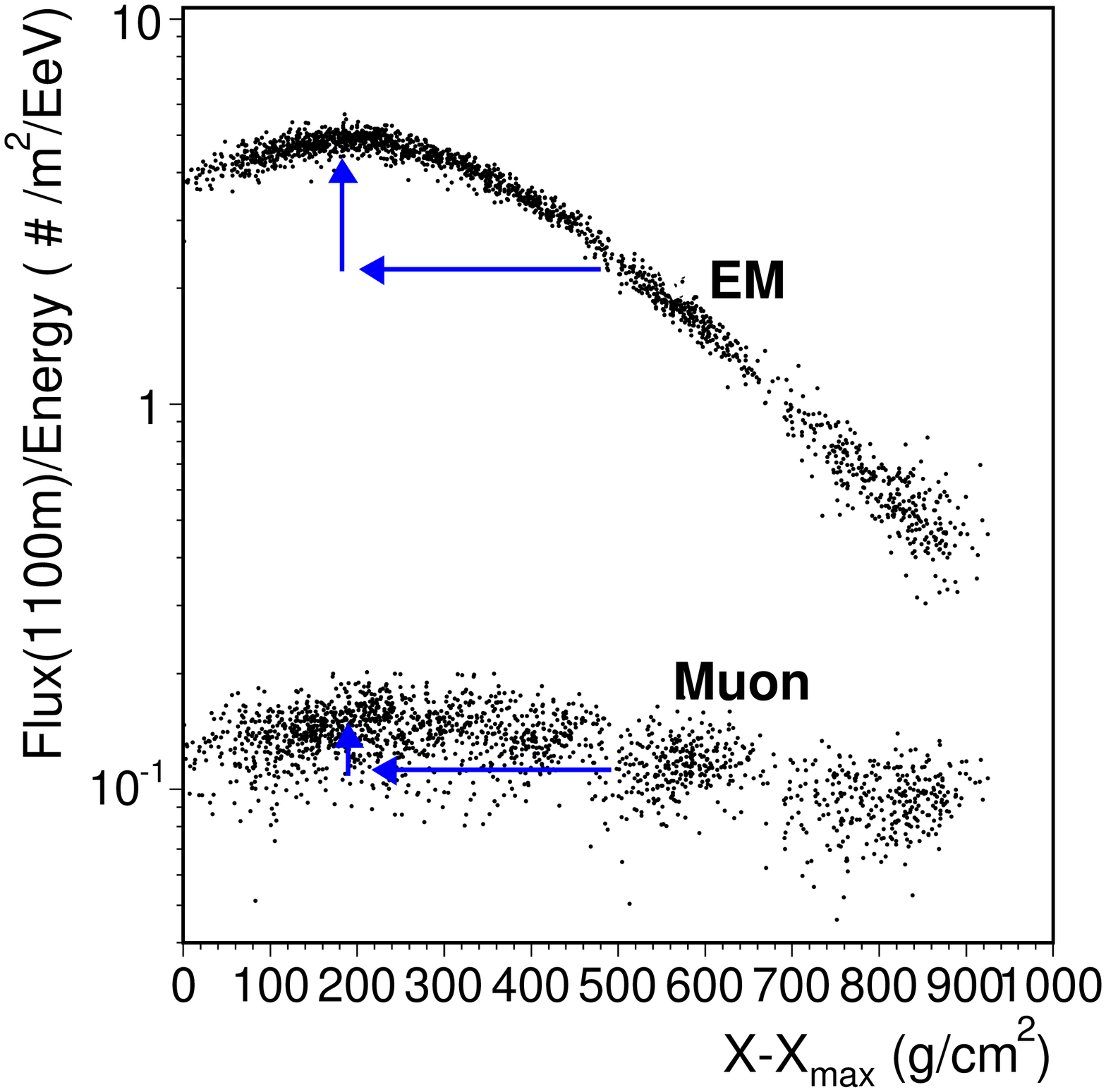}}
\end{minipage}
\caption{\label{F:groundflux} (left) The simulated particle number flux at 500 m core distance, as might be measured by AGASA.  The flux is plotted versus the remaining depth between $X_{max}$ and the depth of the detector.  A deep shower can deposit $\sim$3 times the expected flux from ordinary showers.  This induces a large overestimate of the primary energy. (right) The simulated particle number flux at 1100 m core distance, as might be measured by Auger.  While a deep shower may deposit a much larger-than-expected EM flux, the muon flux would be roughly the same.  The flatness of the muon flux combined with Auger's enhanced sensitivity to muons tempers the tendency to overestimate the energy of deep showers.  }
\end{figure}

In Figure~\ref{F:groundflux}(left) we plot the simulated electromagnetic particle flux $\rho(500)$ as a function of the depth $X-X_{max}$ of the remaining atmosphere that the shower must traverse to get from $X_{max}$ to the detection position.  (For technical reasons we use 500 m instead of 600 m, but the plots are qualitatively the same.)   The AIRES shower simulation package \cite{Sciutto:AIRES} is used with proton primaries and the QGSJET01 \cite{qgsjet} hadronic interaction model.~\footnote{The simulation is taken from a shower library generated by Sergio Sciutto, using the Fermilab Fixed Target computing farm.}  AGASA's thin unshielded scintillators measure primarily the electromagnetic ($e^\pm,\gamma$) particle number flux, which dominates over the muon flux for ordinary near-vertical showers.  The electromagnetic (EM) flux attenuates as expected for large depths $X-X_{max}$, and the peak position is shifted away from $X=X_{max}$ precisely because of the effect described above of the finite transverse distance to the detection position.  AGASA uses the data sample with zenith angles $\theta<$45$^\circ$ for their energy spectrum analysis, which corresponds approximately to $X-X_{max}<$ 500 g/cm$^2$.  This event selection corresponds to the region in the plot where the electromagnetic flux curve is relatively flat, and hence the energy measurement is relatively insensitive to the shower-to-shower fluctuations in $X_{max}$.  The EM flux curve may be thought of as a correction factor for the expected attenuation of the EM shower, based on the distance that the shower is expected to have propagated before reaching ground.  This correction factor is applied when converting the measured $\rho(600)$ to the inferred primary energy.

Now imagine if an exotic cosmic ray enters the atmosphere and penetrates much more deeply than an ordinary hadronic cosmic ray would have penetrated, resulting in an $X_{max}$ located much closer to the detection position.  Then the attenuation correction factor can be greatly overestimated, and hence the energy is also overestimated.  For example, if an ordinary shower at $\theta = 45^\circ$ is expected to reach its maximum at $X-X_{max}=$ 500 g/cm$^2$, but actually reaches maximum much closer to the ground at $X-X_{max}=$150 g/cm$^2$, then the energy is overestimated by a factor of $\rho_{true}/\rho_{expected}\sim$ 3.  It is therefore possible that the super-GZK events reported by AGASA have their energies greatly overestimated if the typical $X_{max}$ of those events is much larger than the range of $X_{max}$ values expected from ordinary hadronic showers.  For this argument we have assumed for simplicity that the longitudinal development profile of an exotic shower is similar to that of an ordinary shower, other than being displaced towards larger depth.  Large distortions in the shape of the longitudinal profile could yield different model-dependent predictions for the energy bias.  

The systematic energy bias would tend to increase with zenith angle $\theta$, but AGASA's 11 highest energy events appear to be evenly distributed in the acceptance-weighted quantity $sin^2\theta$.  However, the prediction of a peaked distribution in $sin^2\theta$ may be relaxed if the exotic particles originate from a finite number of non-uniformly distributed sources in the sky.  Also, if the ``super-GZK'' flux has a mixed composition of ordinary and exotic cosmic rays, then different source distributions and/or different systematic biases in the acceptance and energy reconstruction of each component may conspire to smooth out distortions in the zenith angle distributions.  For example, the tendency of exotic sub-GZK cosmic rays to be peaked at large $\theta$ may be countered by a tendency for ordinary super-GZK cosmic rays to be peaked at small $\theta$.

In principle, the hypothesis that the super-GZK events are not super-GZK after all, but simply have $X_{max}$ positions very close to the ground can be easily tested with the AGASA raw data by measuring the curvature of the shower front in these events using the delay of the trigger time as a function of the core distance of the detector.  The radius of curvature is roughly related to the distance of $X_{max}$ from the ground, and is typically of the order of $\sim 15$ km.  With timing measurements up to transverse distances of only 2 km as well as a partial degeneracy between the curvature measurement and the arrival direction measurement, it is difficult to make a precise determination of the curvature.  However, it should still be possible to distinguish deeply penetrating showers with $X_{max}$ very close ($<$2 km ) to the ground from ordinary showers with $X_{max}$ far away.  It is also worth noting that exotic particles which are less deeply penetrating than typical hadronic cosmic rays would have their energies underestimated by AGASA and hence would be buried underneath the dominant $1/E^3$ spectrum.  It is unlikely that AGASA would have a curvature resolution precise enough to distinguish these kinds of exotic showers from ordinary showers.  Similarly, underestimation of the energy may occur if the showers are so deeply penetrating that they have not yet fully developed before hitting the ground.  The shape of the lateral distribution of measured particle fluxes may therefore also provide useful information relating to the scale of the transverse development of the showers.

\section {Exotic particles as viewed by HiRes}

The HiRes1 and HiRes2 monocular fluorescence data show evidence of a suppression of the cosmic ray spectrum at GZK energies.  However, as we shall discuss below, the analysis procedures for each detector may bias the event selection towards ordinary hadronic primaries.  If the event selection criteria efficiently remove the small fraction of exotic events, then their evidence for a GZK suppression correctly implies that there is no new physics in the interactions of protons and ordinary nuclei with the cosmic microwave background.  In particular, there would be no need to invoke exotic models such as Lorentz violation in order to suppress the GZK interactions.  The small discrepancy between the AGASA spectrum and the HiRes spectra could then be viewed as an indication of new non-GZK physics.  We can consider two possibilities.  First, the AGASA events really have super-GZK energies but are rejected by the HiRes event selection.  Second, the AGASA events are really mismeasured sub-GZK events which may or may not be rejected or mismeasured by the HiRes analysis procedures.  Even if they do make it into the HiRes spectrum sample, since they comprise a small flux of sub-GZK exotic cosmic rays, it is likely that they would be hidden beneath the much larger flux of ordinary cosmic rays.

As a cosmic ray shower traverses the sky, the large flux of charged particles simultaneously deposits ionization energy in the atmosphere, and excites nitrogen molecules which then emit ultraviolet fluorescence light.   The HiRes telescopes image the fluorescence light emitted by nitrogen molecules onto PMT cameras and produce a series of measured pulseheights $S$ as a function of viewing angle $\chi$ within the shower-detector-plane (SDP) and time.  If the trajectory of the shower can be reconstructed using the time vs $\chi$ data, then it is possible to make a mapping of $\chi$ onto penetration depth $X$.  The plot of the PMT pulseheights $S$ versus $X$ may then be corrected for detector efficiencies, atmospheric attenuation, and fluorescence yield in order to obtain the longitudinal profile of the number of minimum ionizing shower particles versus $X$.  An universal Gaisser-Hillas function \cite{gaisserhillas} is fitted to this data in order to account for the the portions of the longitudinal profile which are outside of the field of view of the telescope.  The Gaisser-Hillas function shown in equation~\ref{gheqn} has four parameters:  the normalization $N_{max}$, the first interaction depth $X_0$, the depth of shower maximum $X_{max}$, and an attenuation length $\lambda$.  The integral of of this function multiplied by the average $dE/dX$ then gives the total ionization loss which is proportional to the initial cosmic ray energy.

\begin{eqnarray} 
N(X) = N_{max} \left(
\frac{X-X_0}{X_{max}-X_0} \right)^\frac{X_{max}-X_0}{\lambda}
e^\frac{X_{max}-X}{\lambda}\,, 
\label{gheqn} 
\end{eqnarray}

\begin{figure}
\resizebox{12cm}{!}{\includegraphics{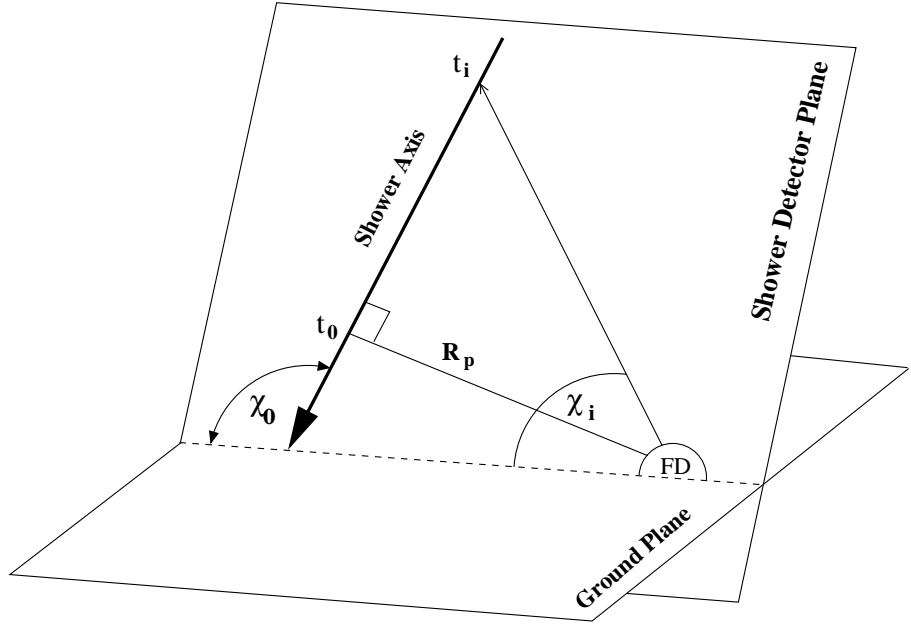}} 
\caption{\label{F:fdparms} Figure defining observables within the shower-detector-plane (SDP).  $\chi_0$ is the shower angle w.r.t. the ground.  $R_p$ is the impact parameter, and $t_0$ is the time at the point of closest approach to the telescope.  These three parameters may be inferred from the dependence of the pixel trigger time $t_i$ on the pixel viewing angle $\chi_i$. 
}
\end{figure}

The bulk of the HiRes data come from the HiRes1 detector which has a field of view of 14$^\circ$ in elevation angle.  Because of the rather limited field of view, the measured tracklengths are too short to reliably reconstruct the shower trajectory from the time-vs-$\chi$ data.  Because the geometry fit extracts three parameters: the angle $\chi_0$ of the shower with respect to the ground, the impact parameter $R_p$ of the shower to the telescope, and the time of the shower $t_0$ (see figure~\ref{F:fdparms}), the track in the camera must be long enough to be able to extract at least the zero-th, first, and second derivatives of the time-vs-$\chi$ curve in order to reconstruct the trajectory.   Since $d^2\chi/dt^2$ is typically very small, it is difficult to measure this parameter in short tracks and the large resulting uncertainty in the shower geometry gives large uncertainties in the mapping of $\chi$ to $X$.

Instead, the HiRes1 data is analyzed using a profile-constrained-fit (PCF) in which the longitudinal profile and the shower geometry are simultaneously reconstructed \cite{Abbasi:2002s}.  In this fit, the parameter $X_0$ is fixed to be 0 to reflect the fact that hadronic showers start near the edge of the atmosphere.  The attenuation parameter $\lambda$ is fixed to be 70 g/cm$^2$, and the position of shower maximum $X_{max}$ is allowed to vary among a set of discrete values between 680-900 g/cm$^2$, the range of typical $X_{max}$ values measured by the Fly's Eye stereo experiment \cite{Bird:1993c}.  At each trial value of $X_{max}$, a two-parameter fit is performed to extract the normalization parameter $N_{max}$ and the shower angle $\chi_0$.  The $X_{max}$ which gives the best overall fit is selected, and the fitted function is then integrated to extract the shower energy. 

\begin{figure}
\begin{minipage}[t]{0.49\textwidth}
\mbox{}\\
\centerline{\includegraphics[width=\textwidth]{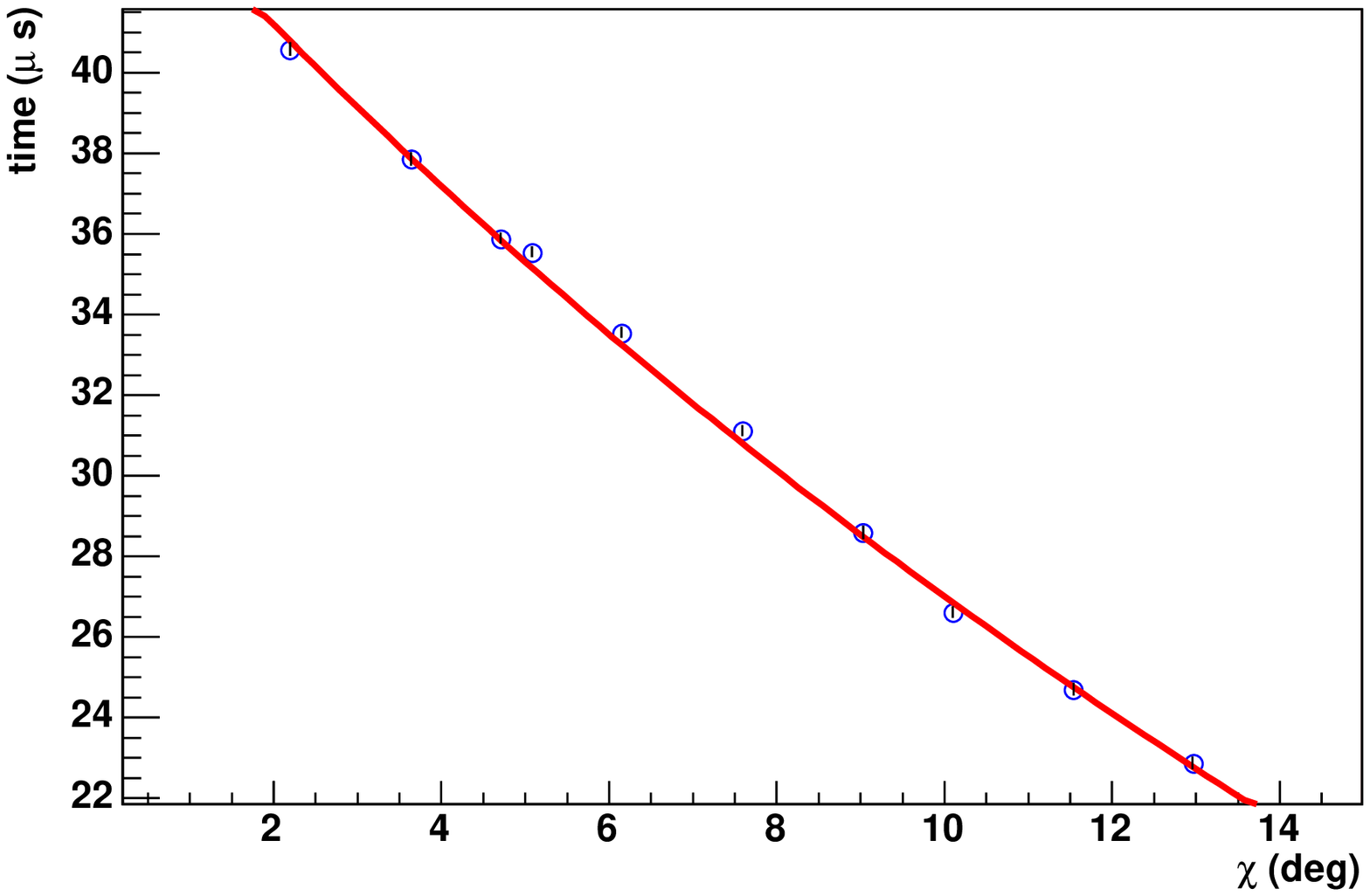}}
\end{minipage}
\hfill
\begin{minipage}[t]{0.49\textwidth}
\mbox{}\\
\centerline{\includegraphics[width=\textwidth]{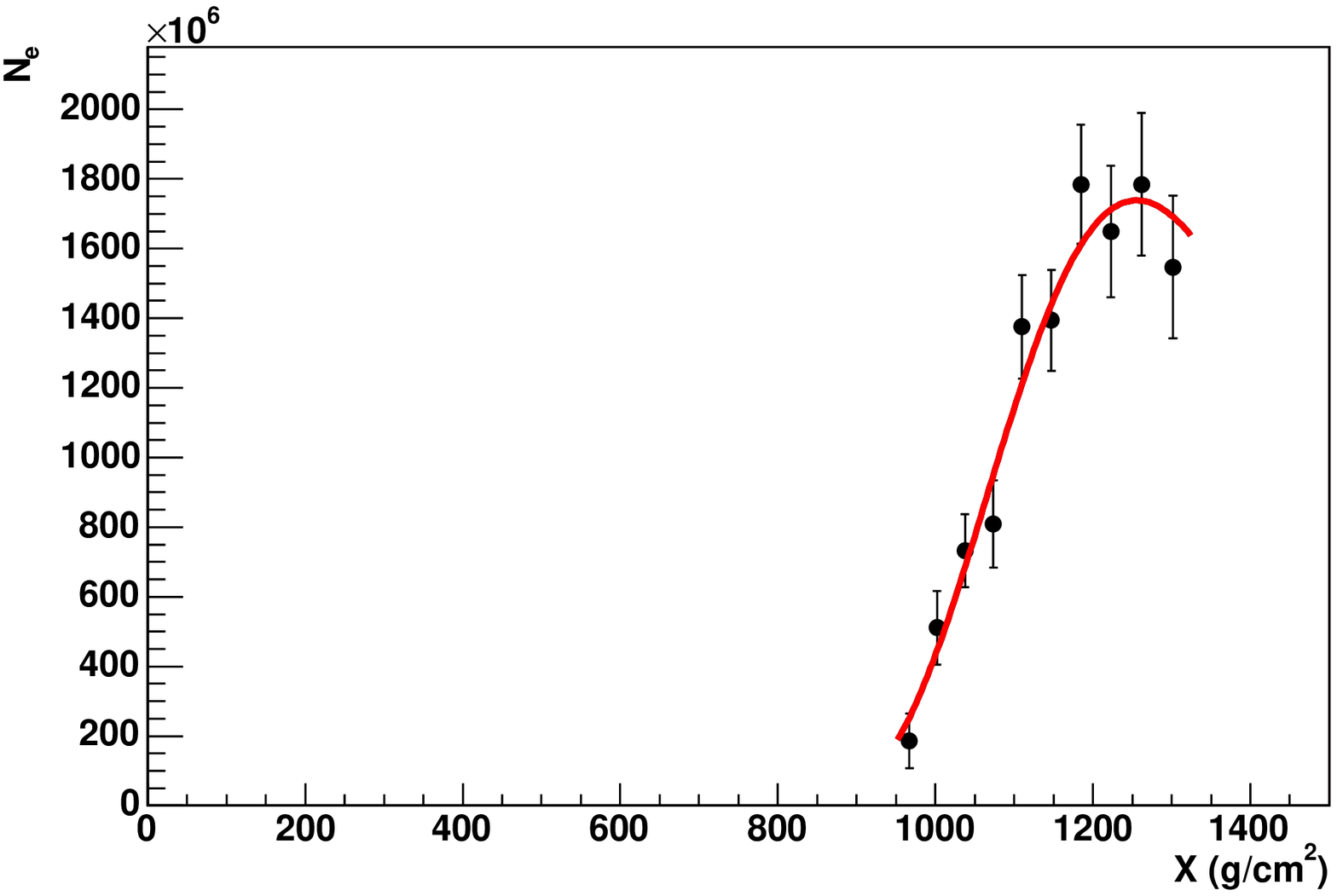}}
\end{minipage}
\begin{minipage}[t]{0.49\textwidth}
\mbox{}\\
\centerline{\includegraphics[width=\textwidth]{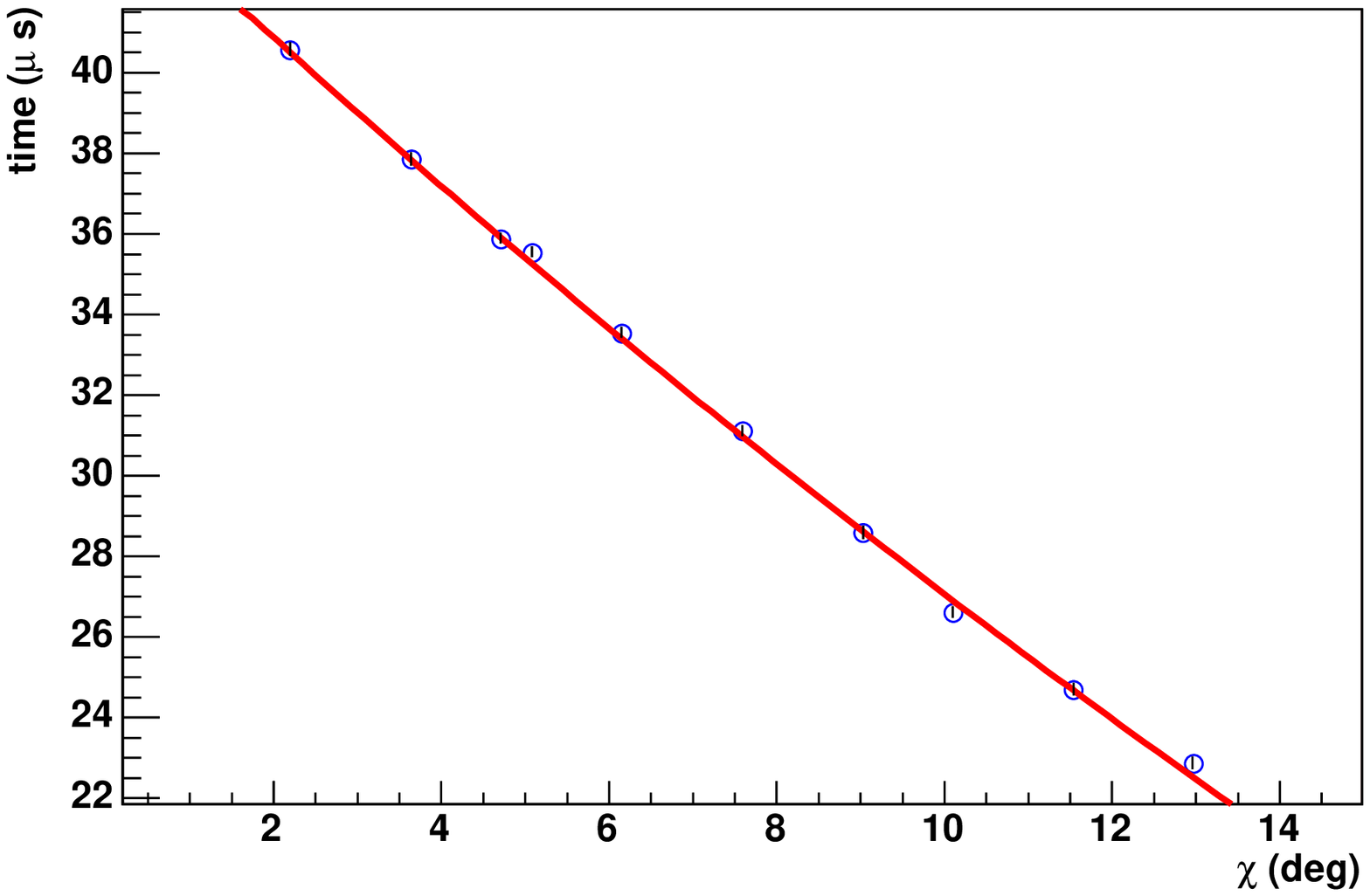}}
\end{minipage}
\hfill
\begin{minipage}[t]{0.49\textwidth}
\mbox{}\\
\centerline{\includegraphics[width=\textwidth]{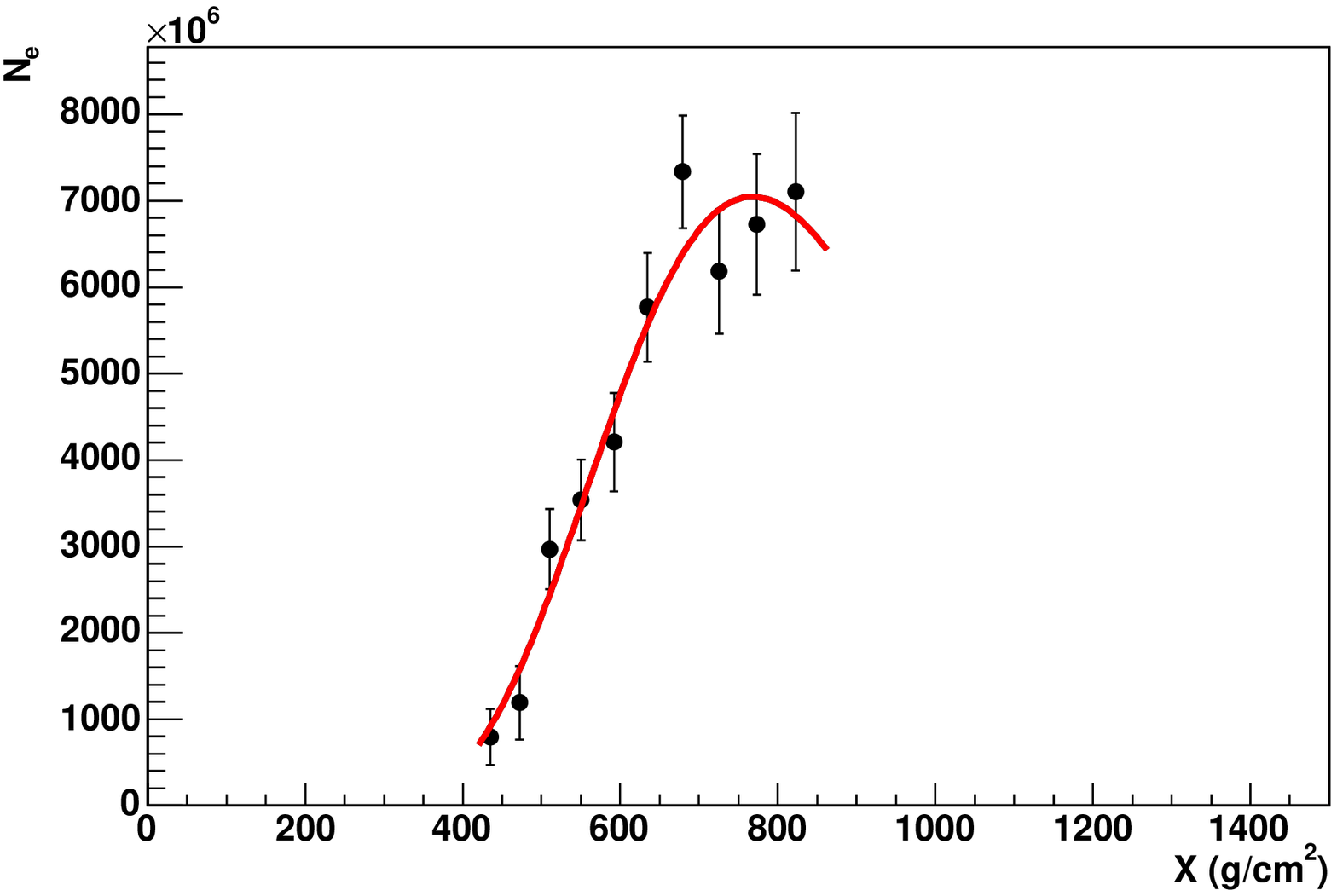}}
\end{minipage}
\caption{\label{F:shorttrack} The time vs viewing angle plots (left) and the inferred longitudinal profiles (right) for the $\chi_0 = 140^\circ$ (top) and $\chi_0 = 110^\circ$ (bottom) interpretations of data from a short track.}
\end{figure}

 The HiRes PCF procedure is manifestly designed to reconstruct ordinary hadronic showers induced by cosmic rays with typical hadronic scattering cross-sections which yield $X_{max}$ values in the selected range.  If the $X_{max}$ of an exotic shower is deeper than this allowed range, then the geometry of the shower will be misreconstructed, and either the event will fail to pass some other event selection cut, or the event will make it into the spectrum analysis sample but with an incorrect energy.  To illustrate the possible effect, we take as an example a simulated deep shower which is viewed as a short track in HiRes1.  In figure ~\ref{F:shorttrack}, we show the reconstructed longitudinal profile for two fixed values of the shower angle $\chi_0$, each of which is consistent with the timing data.  The timing errors are modelled as 10$\%$ of the PMT traversal time, as suggested in \cite{Wilkinson:1998}.  In the first case, the shower is assumed to be more inclined towards the telescope, and the shower profile is truncated at 1300 g/cm$^2$ as it disappears below the field of view of the telescope.  In the second case, the shower is assumed to be more vertical, and the remapping of viewing angle $\chi$ to depth $X$ causes the whole profile shift to smaller depth.  In either case, the shape of the profile looks reasonable, and the pixel timing data is consistent with the hypothesis.  This example shows that at least in some cases, monocular fluorescence data can be consistent with either a deep or shallow shower interpretation, and the HiRes PCF reconstruction would be biased towards the possibly incorrect shallow shower interpretation.

An identical analysis is performed in the report of the 320 EeV Fly's Eye event \cite{Bird:1994mp}, the highest energy event ever recorded.  In Fly's Eye, as well as in HiRes1, the signal integration electronics do not permit a precise timing resolution because a PMT could have triggered at almost any time during the traversal of the shower image across the PMT.  The timing error, defined as the time difference between the trigger time and the time at which the shower image passed the center of the PMT, is very difficult to measure or estimate.  In practice, the timing errors of both experiments are arbitrarily rescaled to give equal weight to the timing fit and the profile fit.  The statistical errors in all measured parameters are dominated by the timing uncertainties which give rise to uncertainties in the event geometry.  To attempt to reduce the energy of the Fly's Eye event, the authors of \cite{Bird:1994mp} tilt the shower axis to reduce the value of the impact parameter $R_p$, thus moving the light source closer to the telescope.  The apparent brightness at the camera then corresponds to a lower energy source.  They then reject this solution because, due to the larger slant depth of the tilted shower axis, the resulting $X_{max}=$1335 g/cm$^2$ and $X_0>$ 550 g/cm$^2$ are ``implausibly high'' for a shower induced by protons, nuclei, or gamma rays.  Furthermore, the $\chi^2$ value for the lower energy hypothesis is a factor of six larger, but since the timing errors are rescaled, the $\chi^2$ difference is not necessarily meaningful.  In our opinion, a more serious problem is that the width of the longitudinal profile becomes greatly reduced, and perhaps inconsistent with the idea of a reduced nucleon interaction cross-section. Nevertheless, this example illustrates that even for long tracks there can be large reconstruction errors if the timing uncertainties are large.  Also, an analysis bias towards ordinary shallow showers is already manifest in this early publication.

\begin{figure}
\resizebox{16cm}{!}{\includegraphics{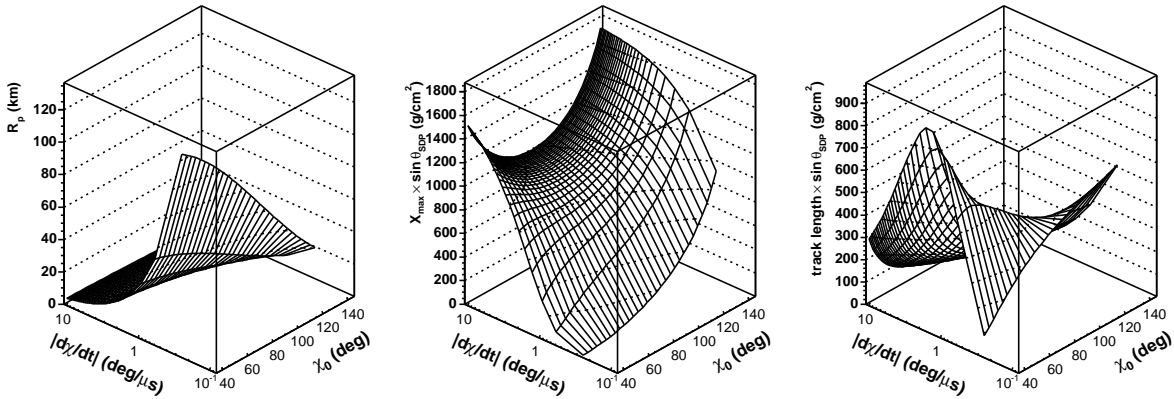}}
\caption{\label{F:analytic} The impact parameter $R_p$, the depth of shower maximum $X_{max}$, and the depth interval $\Delta X$ of a $10^\circ$ track segment, all plotted versus the measured $|d\chi/dt|$ and the postulated $\chi_0$.  A mean viewing angle of $\chi_m = 10^\circ$ is assumed. 
}
\end{figure}

To investigate the PCF analysis in more detail, we compute the analytic functions describing $R_p$ and $X_{max}$ as a function of the mean viewing angle $\chi_m$ and the average measured angular velocity $d\chi/dt$ in a track, and the assumed $\chi_0$.   The dependence on $d^2\chi/dt^2$ is suppressed because it cannot be measured in short tracks.  It is also assumed that for well-measured showers, $X_{max}$ will be viewed at an elevation angle of approximately $\chi_m$ so that a maximum can be clearly seen in the measured profile.  Since $d^2\chi/dt^2$ is very small, $d\chi/dt$ is well measured even for short tracks, and typically takes values between 0.1-10$^\circ /\mu$s.  To evaluate the scaling of the width of the longitudinal profile, we also compute $\Delta X(\chi_m,d\chi/dt,\chi_0)$, the interval in depth corresponding to a $\chi_1-\chi_2=$10$^\circ$ interval in viewing angle.  The standard U.S. atmospheric model of Linsley \cite{linsleyatm} is used to compute the vertical air density profile for the function $D(x)$ which converts from height above the telescope altitude to vertical atmospheric depth.  The formulae are shown in equations~\ref{rpeqn}-~\ref{dxeqn}.  In particular, equation~\ref{rpeqn} describes the 1-parameter degeneracy in the geometry between $\chi_0$ and $R_p$ when $d^2\chi/dt^2$ cannot be measured.  In figure~\ref{F:analytic}, we plot these functions versus $d\chi/dt$ and $\chi_0$, with $\chi_m$ fixed to 10$^\circ$.  The plots also use a vertical shower detector plane (SDP) whose normal vector has a zenith angle $\theta_{SDP}=$90$^\circ$, but should be valid for small deviations from vertical.  

\begin{eqnarray}
R_p = 2 c \cdot \frac{dt}{d\chi} \cdot \cos^2\left(\frac{\chi_0-\chi_m}{2}\right) 
\label{rpeqn}
\end{eqnarray}

\begin{eqnarray}
X_{max} = \frac{D(R_p \cdot (\cot{(\chi_0-\chi_m)} - \cot{\chi_0}) \cdot \sin{\chi_0} \cdot \sin{\theta_{SDP}})}{\sin{\chi_0} \cdot \sin{\theta_{SDP}}}
\label{xmaxeqn}
\end{eqnarray}

\begin{align} \Delta X = &\frac{D(R_p \cdot (\cot{(\chi_0-\chi_1)} - \cot{\chi_0}) \cdot \sin{\chi_0} \cdot \sin{\theta_{SDP}})}{\sin{\chi_0} \cdot \sin{\theta_{SDP}}} \notag \\ 
&- \frac{D(R_p \cdot (\cot{(\chi_0-\chi_2)} - \cot{\chi_0}) \cdot \sin{\chi_0} \cdot \sin{\theta_{SDP}})}{\sin{\chi_0} \cdot \sin{\theta_{SDP}}}
\label{dxeqn}
\end{align}

The $R_p$ plot indicates that at any fixed $d\chi/dt$, as $\chi_0$ increases, the shower is tilted towards the telescope and $R_p$ decreases.  As $R_p$ decreases, the atmospheric attenuation correction becomes smaller, and so for a fixed signal pulseheight the energy estimate decreases.  The $X_{max}$ plot indicates two distinctive trends.  For showers passing near the telescopes, the angular velocity $|d\chi/dt|$ is large.  As showers are tilted from being inclined away from telescope towards being being inclined towards the telescope, the $X_{max}$ position remains in a region of nearly constant density in the lower atmosphere.  Therefore, $X_{max}$ closely follows the $\sec{\theta}$ slant depth of the assumed geometry.  For showers far away from the telescope, $|d\chi/dt|$ is small.  In this case, if the geometry is inclined away from the telescope to lower chi, then the viewed portion of the shower at very large $R_p$ is placed in the low density upper atmosphere.  In this case, the measured $X_{max}$ which is assumed to be in the field of view, is very small.  The effect of placing the track in the lower or upper atmosphere is also visible in the tracklength plot.  For nearby showers where the track is viewed in a constant density region, the tracklength is simply geometrical and decreases monotonically as the shower is tilted towards the telescope, due mainly to the $R_p$ lever arm suppression.   For far away showers, although the geometric distance intervals increase with $R_p$, the corresponding depth intervals $\Delta X$ are greatly suppressed for geometries at small $\chi_0$, tilted away from the telescope.  

These plots indicate that for short tracks with finite timing resolution, it is easy to get order unity errors in the $X_{max}$ by assuming that the $X_{max}$ of exotic showers must be in the interval allowed by the PCF procedure.   For exotic deep showers, the PCF procedure corresponds to tilting the showers towards vertical ($\chi_0$=90$^\circ$) in order to reduce the slant depth.  For nearby showers, because both $R_p$ and $\Delta X$ are monotonically decreasing with $\chi_0$, the measured energy as well as the measured tracklength can either increase or decrease as $\chi_0$ approaches $90 ^\circ$ from either side.  For far away showers, $\chi_0$ must be reduced (tilting the shower axis away from the telescope) in order to force the PCF fit.  In this case, $R_p$ increases and so the energy is overestimated.  On the other hand, $\Delta X$ decreases, and the measured profile might not have the width $\lambda=$70 g/cm$^2$ required by the HiRes Gaisser-Hillas fit or by the event selection cuts.  In either case, unless the energies are greatly overestimated, it is unlikely that such exotic events would be obviously visible above the flux of ordinary non-exotic cosmic rays.  Further analysis would require a more detailed simulation of the HiRes1 detector which is beyond the scope of this work.  It is worth mentioning however that in the preliminary report \cite{bird:1995}, the estimated uncertainty $\sigma(R_p)/R_p \sim$1 for typical 10$^\circ$ tracks in the HiRes1 detector.  This corresponds to a 50$^\circ$ uncertainty in $\chi_0$ in our plots which gives large uncertainties in $X_{max}$.  Since energy scales roughly as $R_p$, it is also uncertain by a factor of two.

 The HiRes2 FADC data comprise a much smaller portion of the total HiRes dataset, but the detector has the advantage of having a 28$^\circ$ field of view in elevation, and much more precise timing resolution \cite{Abbasi:2005ni}.  For the set of long tracks selected in the HiRes2 analysis, a timing fit can be performed with a resolution of $\sigma_{\chi_0}\sim$5$^\circ$.  In this case, much of the uncertainty in the conversion of viewing angle to penetration depth can be removed.  However, in the profile fitting procedure, HiRes2 still fixes the parameter $X_0=$ 0 g/cm$^2$, again reflecting the assumption that cosmic rays are protons or nuclei.  Showers with deep profiles which begin at large values of $X_0$ would not be well-fitted by this procedure, and the resulting large chi-squared values might cause such events to be rejected at the event selection stage.  Similar remarks apply to the HiRes Stereo dataset \cite{Abbasi:2004nz}.  Low statistics combined with the possible fitting bias make it seem likely that exotic events may have eluded detection, even in these higher quality datasets.

\section {Exotic particles as viewed by Auger}

The Pierre Auger Observatory \cite{Abraham:2004dt} consists of both a surface detector array and multiple fluorescence telescopes looking inward over the array.  Because it employs both detector technologies, it offers an exciting opportunity to resolve the degeneracies in the measurements of each individual technique between exotic and ordinary interpretations of the event data.  Since the fluorescence detectors have only a 10$\%$ duty cycle, operating only on clear, dark nights, a high statistics energy spectrum measurement is ideally made with the surface detector data.  To calibrate the surface detector energy measurement, Auger uses simultaneous "hybrid" observations of cosmic ray air showers with events which independently trigger the fluorescence detectors and the surface detectors.  The geometry of the events measured in hybrid mode is very well determined by having both longitudinal and transverse timing constraints \cite{Bonifazi:2005ns}.  By plotting the well-measured fluorescence cosmic ray energy versus a ground flux normalization parameter $S_{38}$ (the attenuation-corrected signal at 1000 km core distance), Auger obtains a roughly linear relationship between the ground parameter and the calorimetric fluorescence energy \cite{Sommers:2005vs}.  This energy formula derived from the hybrid data is then applied to the entire Auger surface detector data set to obtain an energy spectrum with higher statistics.  The hybrid calibration should give a reliable energy measurement for typical cosmic rays, since the calibration reflects the average behavior of cosmic ray air showers seen in the hybrid dataset.  In this data-driven approach, the cosmic rays are not assumed a priori to be protons or nuclei.  However, there is an implicit assumption that all high energy cosmic rays produce air showers with similar characteristics which are well-described by their average behavior.  

In principle, deeply penetrating showers would lie well away from the fitted calibration curve because while the fluorescence measurements would remain calorimetric, the ground flux would be greatly enhanced.  However, if the fraction of exotic events is small, for example of order ten ``super-GZK'' events in the AGASA dataset, then there should be only of order one such event in the Auger hybrid data, and perhaps even fewer due to the very strict quality criteria applied when selecting the calibration events.  Furthermore, as is the case with the HiRes analyses, there may be hidden biases in the fluorescence $dE/dX$ profile reconstruction procedure which would tend to reject shower profiles which do not resemble typical hadronic showers.  For example, the standard Auger analysis also fixes $X_0$ to a small value when performing the fit, and so large chi-squared values may cause the events to be rejected.  Another possibility is that if the deep showers typically hit the ground before reaching shower maximum, then these events would still trigger the surface detector, but be rejected by the fluorescence analysis because the profile cannot be reliably extrapolated below the field of view of the telescope.  

\begin{figure}[!t]
\begin{minipage}[t]{0.48\textwidth}
\mbox{}\\
\centerline{\includegraphics[width=\textwidth]{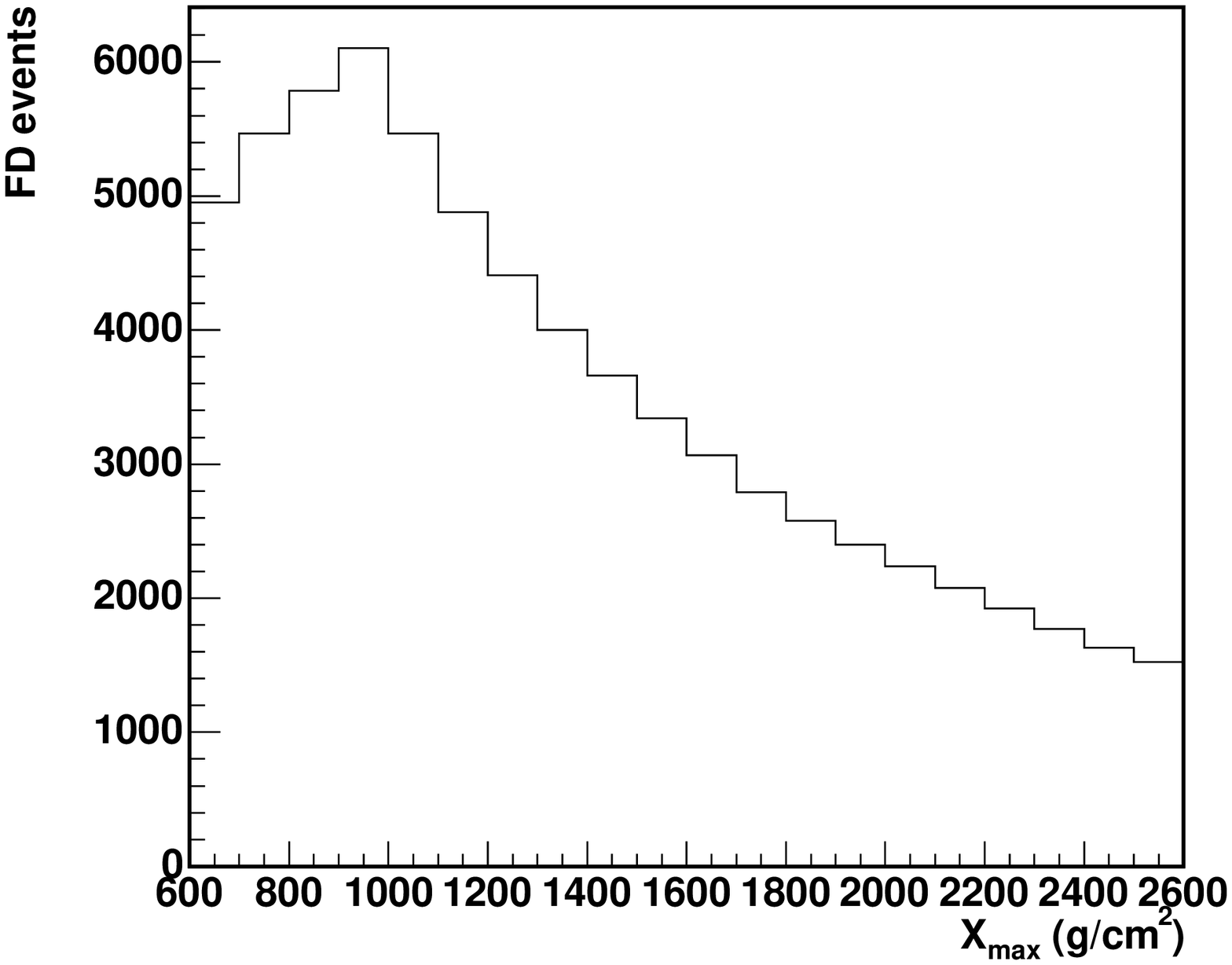}}
\end{minipage}
\hfill
\begin{minipage}[t]{0.48\textwidth}
\mbox{}\\
\centerline{\includegraphics[width=\textwidth]{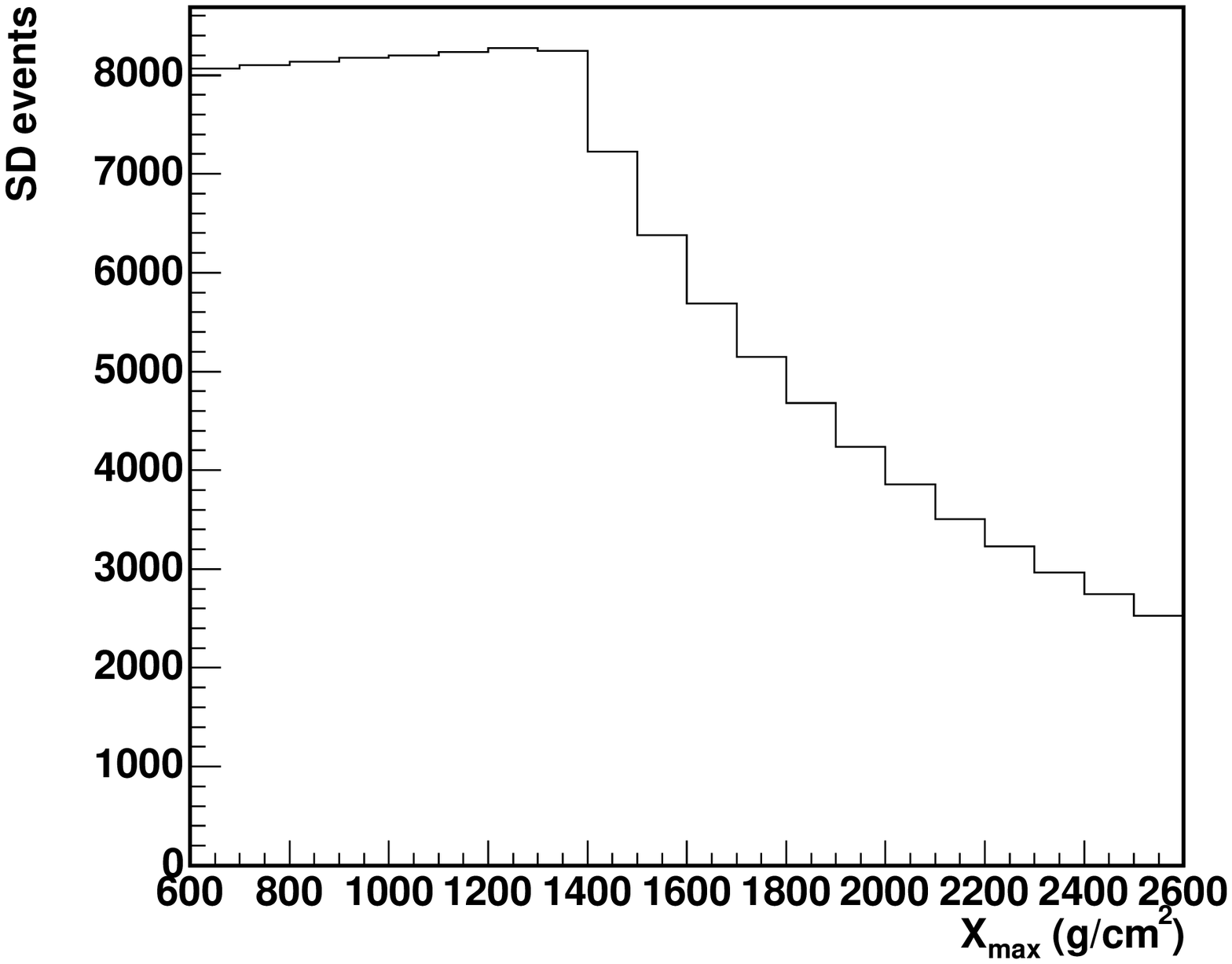}}
\end{minipage}
\caption{\label{F:acceptance}  Fast Monte Carlo simulation of the relative acceptances vs $X_{max}$ of the Auger fluorescence detector (left) and surface detector array (right). }
\end{figure}

There are also different geometric acceptances for the fluorescence telescopes and the surface detector array.  For large $X_{max}>$ 1000 g/cm$^2$, the fluorescence acceptance as a function of $X_{max}$ falls as $1/X_{max}^2$ because large $X_{max}$ values correspond to a large zenith angles.  For any viewing volume, the range of zenith angles corresponding to a fixed $X_{max}$ binsize shrinks as the zenith angle (and hence $X_{max}$) increases.      A fast Monte Carlo simulation of the relative acceptances as a function of $X_{max}$ of the fluorescence telescopes and the surface detector array is shown in figure~\ref{F:acceptance}.  The $10^{19}$ eV events are simulated with flat distributions in core position, in solid angle, and in $X_{max}$, and the resulting $X_{max}$ distribution of events which triggered and passed typical event quality cuts is histogrammed.   The fluorescence simulation follows the procedure described in \cite{Prado:2005xj}.  The surface detector simulation assumes the Auger array geometry, and a trigger threshold of three vertical muon flux units at 1 km core distance.  The fluorescence acceptance is suppressed at small $X_{max}$ due to showers being above the field of view of the telescope.  As $X_{max}$ increases, the acceptance grows as the showers enter the field of view, and then falls due to the $\sec{\theta}$ divergence of the slant depth.  The surface detector acceptance is flat as a function of $X_{max}$ up until a value of $\sim$1400 g/cm$^2$, at which point it also falls due to geometric effects.  The flatness is believed to be due to the possibility of triggering the surface detector whether $X_{max}$ is high above the ground, or deep underground.  Just as in the case of AGASA, while the surface detector may trigger, the inferred shower energy may be grossly incorrect.  However, since the fluorescence detector and surface detector acceptances have quite different dependences on $X_{max}$, the composition of the hybrid data sample which requires independent triggers by both detectors might not reflect the composition of the larger surface detector data sample.  We also note that the HiRes-AGASA comparison might have a similar problem.  Since HiRes models their telescope acceptance assuming shallower conventional primary particles, they may overestimate the acceptance for deep showers which might be responsible for the AGASA super-GZK signal.  The fluorescence telescope acceptance for deep showers is reduced whether these showers really have super-GZK energies or not, and so even the measured flux of true super-GZK events may be systematically suppressed.

Another subtlety is that Auger's surface detectors are water Cherenkov (WC) detectors of 1.2 m height rather than 5 cm thin scintillators like AGASA's, and are hence much more sensitive to the muon flux.  AGASA scintillators give an approximately equal response to all minimum ionizing particles including $e^\pm$ pairs from gamma conversion, and therefore act as particle counters.  The AGASA signal is therefore dominated by the EM ($e^\pm/\gamma$) flux in the shower which is much larger than the muon flux for ordinary near-vertical showers.  The Auger WC detectors are $\sim 3$ radiation lengths deep however, and serve as calorimeters of EM particles which shower inside the water.  The Cherenkov light yield is proportional to the tracklength and hence to the total energy deposited within the water volume, typically around 5 MeV/particle.  Muons however simply pass through the water with a tracklength determined by the geometry of the track and of the detector.  Using a $dE/dX \sim$ -2 MeV/cm, we can approximate the muonic energy deposit for a vertical track as 240 MeV.  The response of the Auger tanks to muons is therefore a factor of $\sim$50 greater than the response to EM particles.  As a result, the muon flux composes approximately 50$\%$ of the signal flux at 1000 m transverse core distance from which the shower energy is inferred.

Although the Auger WC array would still have a tendency to overestimate the energies of deeply penetrating showers at moderate zenith angles, the effect would be smaller than that of AGASA because the Auger energy is derived from both the EM flux and the muon flux.  While the predicted EM flux of the shower has a steep dependence on the depth between $X_{max}$ and the ground the muonic flux is expected to remain roughly constant.  As an example, the predicted particle fluxes at 1 km core distance as a function of $X-X_{max}$ are shown in figure \ref{F:groundflux}(right).  The flatness of this muon flux curve can be explained as follows.  The individual muons have much larger energies than typical EM particles because once they are created, they do not undergo subsequent radiative scattering.  The loss of muon number flux due to ionization loss leading to muon decay is therefore much smaller than the loss of EM flux due to ionization loss leading to stopping particles.  If we approximate that the muon flux is independent of $X_{max}$ whereas the EM flux increases by a factor of $\sim$2 for a deeply penetrating shower, then we expect that the Auger WC energy derived with equal contributions from both components is overestimated by only 50$\%$.  This can be compared with the much larger overestimation of AGASA energies in the example above which derived from considering only the EM component of the flux.

The above argument implicitly assumes that the deeply penetrating exotic particle produces a shower with similar characteristics as ordinary showers, the only exception being that $X_{max}$ is very large.  We expect that hadronic showers induced by particles scattering off nuclei would exhibit similar generic characteristics regardless of the incoming particle species.  $\sim 90\%$ of the energy would be efficiently transfered to the EM shower and the muonic flux from pion decay would be approximately the same.  If we include the possibility however that the muon flux from exotic showers can be very different from the muon flux in ordinary showers, then there can be no robust prediction of the systematic error in the Auger WC energy measurement.  In fact, it is easy for Auger to underestimate the shower energy.  For example, for pure EM showers initiated by super-GZK photons, the showers are deeply penetrating due to the suppression of bremstrahlung and pair-production by the Landau-Pomeranchuk-Migdal (LPM) effect \cite{Landau:1953um} \cite{Migdal:1956tc}.  The enhancement in the resulting EM ground flux could cause AGASA to overestimate the shower energy over a range of intermediate zenith angles.  However, since the muon flux in a purely EM shower is only a small fraction of the muon flux in a hadronic shower, the signals induced in the Auger WC detectors would be much smaller than the signals induced by a hadronic shower of the same energy, regardless of the $X_{max}$ of the shower.  Previous studies \cite{Risse:2005hi} \cite{Sommers:2005vs} have shown that the energy of the EM shower would be underestimated by $\sim$50$\%$ when using the standard Auger energy conversion formula based on hybrid observations of typical hadronic showers.

\section{Discussion}

In this note, we have described how exotic deep shower events would be treated by the experimental techniques of the AGASA, HiRes, and Auger experiments.  In performing a energy spectrum analysis, each experiment strives to develop an analysis procedure which is valid for the vast majority of cosmic ray events, which are presumed to be ordinary hadrons.  AGASA assumes that cosmic ray air showers look like those predicted by hadronic simulations.  HiRes and Auger assume that the air showers have an $X_{max}$ distribution similar to previously measured distributions when performing the fluorescence reconstruction.  Finally, Auger assumes that all cosmic ray air showers may be characterized with a single energy conversion function calibrated with hybrid data.  Because these spectrum analyses are not specifically tuned to search for a possibly small flux of exotic events, the exotic events would appear as a small amount of noise above the underlying spectrum of ordinary cosmic rays.  

The effect of analyzing deep showers as ordinary showers would seem to be the same for all three experiments--in most cases, the shower energy would be overestimated.  However, the magnitude of the overestimate can be very different for each experiment.  A deep shower inducing a large factor of $\sim$3 overestimate in AGASA would only give a $\sim$50$\%$ overestimate in the Auger surface detectors, due mainly to the flat dependence of the ground muon flux on the shower position.  It is also not unreasonable to expect possible factor of $\sim$2 overestimates in the HiRes1 PCF analysis, if the mis-reconstructed deep showers still pass the event selection cuts.  In a spectrum falling rapidly as the third power of energy, a 25$\%$ fraction of exotic cosmic rays near the GZK energy could easily produce a spectrum with no apparent GZK cutoff if their energies are overestimated on average by a factor of two.  This fraction could be even smaller if the energy systematic error is larger, or if for whatever reason some of the observed high energy events really have do have super-GZK energies.

In HiRes2, HiRes stereo and Auger hybrid, where the geometry is much better reconstructed, there should be no systematic overestimate of the energy due to geometry errors but the geometric aperture based on the average shower $X_{max}$ can easily be overestimated.  Furthermore, the event reconstruction and selection procedures may still cause the exotic events to be rejected.  We have discussed in particular the bias due to fixing the first interaction point $X_0$ in the Gaisser-Hillas fit.  There may be an additional bias from fixing the shower attenuation parameter $\lambda\sim$ 70 g/cm$^2$, or even assuming that the exotic profile shape can be well-described by a Gaisser-Hillas function.  In this paper, we have implicitly assumed that after the first few interactions, all showers develop in the same way and thus yield the approximately same longitudinal profile shape.  However, simulations of showers initiated by exotic heavy hadrons \cite{Albuquerque:1998va} \cite{Berezinsky:2001fy} indicate a tendency to broaden the longitudinal profile due to a lower energy transfer per interaction.  In future studies, the profile shape requirements may be relaxed to perform a more general search.

Specific searches for exotic events have thus far been confined to searches for high energy gammas as predicted by top-down production models.  These searches are tuned using guidance from photon-shower simulations in order to maximize their acceptance for photons, and to reduce systematic bias.  Furthermore, the null results are interpreted within the context of the shower simulations, which may have large systematic uncertainties.  

The AGASA super-GZK signal may be our first indication of new physics in high energy cosmic rays, and the AGASA-HiRes discrepancy may indeed give us some hints on how to identify the new underlying physical processes.  We propose to perform model-independent searches for deep showers which may be a signal for new kinds of exotic particles.  In the case of AGASA, an analysis of the shower front curvature may reveal some outlying events which are positioned close to the ground.  In the case of HiRes1, it is probably difficult to infer any additional information from the data, given that the geometries cannot be independently measured from the pixel timing.  In the cases of HiRes2, HiRes Stereo, and Auger, where the shower geometry is measured better, it appears to be a simple matter to free the parameter $X_0$ in the Gaisser-Hillas fits in order to avoid rejecting showers whose profiles are deeper than expected.  The event selection criteria must also be carefully studied to remove any bias which may be present.  The preliminary study of the acceptance of fluorescence and surface detectors reported here can be refined with more detailed simulations.  The main difficulty is to be able to differentiate between exotic deep showers and ordinary deep showers in the exponential tail of the $X_{max}$ distribution resulting from ``punch-through'' of ordinary cosmic ray primaries.  Given enough statistics, a careful analysis of the tail events may reveal deviations from a pure exponential distribution.  The Auger observatory is producing a high rate of events with well-measured hybrid geometries and longitudinal profiles.  We are optimistic that the deep shower hypothesis can be tested fairly quickly, perhaps even with existing datasets.  

\medskip

{\it Acknowledgments:} We would like to thank Nicolas Busca, Maximo Ave, Tokonatsu Yamamoto, and Dan Hooper, as well as the Auger groups at FNAL and the University of Chicago for interesting discussions. We would also like to thank Sergio Sciutto for generating the AIRES shower library at Fermilab.   This work has been supported by the US Department of Energy under contract No. DE-AC02-76CH03000.

\end{document}